\documentstyle[preprint,cite,aps]{revtex}

\begin{document}
\draft \title{\Large On the Non-Abelian Aharonov Bohm Scattering of
  Spinless Particles} \author{M. Gomes, L. C.
  Malacarne\footnote{Permanent address: Departamento de F\'\i sica,
    Universidade Estadual de Maring\'a - Av. Colombo, 5790 -
    87020-900, Maring\'a -PR, Brazil.} and A. J. da Silva}
\address{Instituto de F\'\i sica, Universidade de S\~ao Paulo\\
  C.P. 66318 - 05315-970, S\~ao Paulo - SP, Brazil} \date{1998}

\maketitle

\begin{abstract}
The Aharonov Bohm scattering for
spinless, isospin 1/2, particles interacting through a non-Abelian
Chern-Simons field is studied. Starting from the relativistic
quantum field theory and using a Coulomb gauge formulation, the
one loop renormalization program is implemented. Through the
introduction of an intermediary cutoff, separating the regions
of high and low integration momentum, the nonrelativistic
limit is derived. The next to leading relativistic approximation
is also determined. In this approach quantum field theory vacuum
polarization effects are automatically incorporated. 
\end{abstract}
\section{INTRODUCTION}
A great deal of interest has been devoted in recent years to the
study of the Aharonov-Bohm  (AB) effect, the scattering of charged particles
by an impenetrable magnetic flux tube\cite{Aharonov}. This situation
was motivated both by potential applications, which come from many
different areas, and also by some conceptual difficulties found in the AB
scattering of spinless particles. In that case, the Born approximation
failed to reproduce the expansion of the exact result and,
furthermore, the second Born approximation turned out to be
divergent\cite{Corinaldesi}. These issues have been investigated
considering as equivalents the AB effect and the scattering of
particles interacting through a Chern-Simons  (CS) field.  In a
nonrelativistic context, it was shown that, up to one loop, that is, the second
Born approximation, agreement
of the perturbative calculation with the expansion of the exact result
could be achieved by introducing an extra quartic self-coupling of the
scalar particles
\cite{Lozano} tuned to eliminate divergences and restore the conformal
invariance of the tree amplitude. 
For the scattering of
two spin up fermions it was verified \cite{Hagen} that an additional
self-interaction was not needed since its role was provided by Pauli's
magnetic term, in accordance with an earlier conjecture \cite{Lozano}.
However, if the fermions had antiparallel spins the
effect of the magnetic interaction canceled and a divergence showed up.

From a more basic standpoint, one can start directly from a
relativistic quantum field theory of charged particles interacting
through a CS field and then appropriately taking a nonrelativistic
limit. Proceeding in this way, purely quantum field theory effects as
vacuum polarization and anomalous magnetic moment are automatically
incorporated. Such a procedure was applied successfully to the 
study of the AB scattering for both spin 0 and spin 1/2
particles\cite{Gomes1,Gomes2}. The calculation was greatly facilitated
by the introduction of an intermediary auxiliary cutoff which, in the
Feynman integrals, separates the regions of high and low energy. 
This allows a direct
simplification of the integrands and it is closely related to the
methods of effective field theories \cite{Weinberg}.
For
the case of spinless or antiparalel spin fermionic particles, it was
found that the low energy part of the amplitude contains a logarithmic
divergence in the limit of very high intermediary cutoff.
Nevertheless, differently from the nonrelativistic calculations
previously mentioned, without any additional hypothesis, the needed
counterterm is automatically afforded by the high energy part of the
contribution. Besides that, terms absent from the direct
nonrelativistic calculation were determined. These new interactions
come from the high energy part of the amplitudes and modify
in an essential way basic properties of the nonrelativistic scattering.

In this work we want to pursue these investigations by considering the
non-Abelian AB effect for spinless particles, which corresponds to the
problem of particles without spin but carrying isotopic spin scattered
by an isotopic magnetic flux tube.  Cosmic strings and black holes are
the more natural applications of this subject\cite{Strings}.

The non-Abelian AB situation was firstly analyzed in a celebrated paper on
nonintegrable phase factors \cite{Yang} and since then it has been
more quantitatively investigated at the quantum mechanical level
\cite{Oh}. Recently, in a direct nonrelativistic approach, the
scattering has been discussed using a non-Abelian Chern-Simons field to
simulate the flux tube with results similar to the ones mentioned
above for the Abelian case \cite{Bergman,Amelino}. Here we want to begin from
a relativistic formulation which, as said before, already embodies
radiative corrections. In this way we will be able to find
next to leading corrections to the results presented in \cite{Bergman}.
 
We will assume that the basic particles carry isotopic spin 1/2. In
the language of the $2+1$ quantum field theory our system is described
by the Lagrangian density \cite{Jackiw}
\begin{eqnarray}
{\cal L} &=& - \Theta \varepsilon ^{\alpha\beta\gamma} tr\left(A_\alpha
\partial_\beta A_\gamma + \frac{2 g}{3}A_\alpha A_\beta A_\gamma \right) 
+  (D_\mu \Phi)^\dagger (D^\mu
\Phi)
\nonumber \\
&-&m^2 \Phi^\dagger \Phi - \frac{\lambda_1}{4} (\Phi^\dagger \Phi)^2 
- \frac{\lambda_2}{4} (\Phi^\dagger T^a\Phi)^2 \; ,
\label{1}
\end{eqnarray}
where $D_\mu= \partial_\mu + g A_\mu$ is the covariant derivative,
$A^\mu=A^{\mu}{_a}T^a$ and $T_a$  are the generators of the group SU(2).   
As we shall see,
up to one loop, the leading contributions correctly reproduce the
nonrelativistic results whereas the next to leading contributions are
new corrections to the nonrelativistic calculation. As we are mainly
interested in the nonrelativistic limit, in this work we will employ a
strict Coulomb gauge.

Our work is organized such that in section \ref{section1} the nonrelativistic  
theory is briefly considered. We do that not only to fix our notation but
also to make easier the comparison with the results of the nonrelativistic
limit of (\ref{1}). In section \ref{section2}, after discussing the one
loop renormalization program for the relativistic model, we analyze the
two body scattering up to next to the leading nonrelativistic approximation.
Final comments and a discussion of our results are presented at the end of that section.

\section{The nonrelativistic model}\label{section1}

In this section we want to summarize the results of the non-Abelian AB
scattering of spinless particles. To facilitate comparison, we shall
use essentially the same notation as in \cite{Bergman}, but to keep
contact with the results to be derived in the next section, we will
employ a cutoff to regularize the spatial part of the loop integrals
instead of dimensional regularization as it was done in that
reference. The Lagrangian density which specifies the model is
\begin{eqnarray}
{\cal L}_{NR} &=& - \Theta \varepsilon ^{\alpha\beta\gamma} tr\left(A_\alpha
\partial_\beta A_\gamma + \frac{2 g}{3}A_\alpha A_\beta A_\gamma \right) 
+ i \Phi^{\dagger}D_t \Phi - \frac{1}{2 m} ({\bf D} \Phi)^\dagger 
({\bf D} \Phi)
\nonumber \\
&-& \frac{1}{4} \Phi^\dagger_{n^\prime} \Phi^\dagger_{m^\prime} 
C_{n^\prime m^\prime n m}
\Phi_{m}\Phi_{m} - \frac{1}{\xi} tr ({\bf \nabla} {\bf A})^2 - \eta^{* a} 
({\bf \delta_{a b}\nabla}^2 + g f_{abc}
{\bf A^c}. {\bf \nabla}) \eta^b \; ,
\label{2}
\end{eqnarray} 
\noindent
where $\Phi$ is an $n$ component complex field belonging to 
the fundamental representation of the $SU(n)$ group. 
 The generators
of the Lie algebra of $SU(n)$, denoted by  $T^a$ satisfy
\begin{equation}
[T^a, \,T^b]= f_{abc}T^c \label{2a} 
\end{equation}
and are normalized such that
\begin{equation}\label{3}
tr\left( T^a T^b\right) = -\frac12 \delta^{ab}\; .\label{2b}
\end{equation}
\noindent
With respect the constant matrix $C_{n^\prime m^\prime nm}$ we just
assume that it has the most general form compatible with the invariance
of the action under the SU(n) transformations \cite{Bergman}. 

We will use a graphical notation where the CS field, the matter field and
the ghost field propagators are represented by wavy, continuous and dashed
lines respectively. In the Coulomb gauge, obtained by letting $\xi 
\rightarrow 0$, the analytic expressions for these propagators are:

CS field propagator:
\begin{equation}
D^{\mu\nu}_{ba}(k) =D^{\mu\nu}(k) \delta_{ba} 
=\frac{1}{\Theta}\varepsilon^{\mu\nu\lambda}\frac{\bar{k}_{\lambda}}
{{\bf k}^2}\;\delta_{ba}\label{2c},
\end{equation}
where $\bar k^\mu = (0,\vec k)$.

Matter field propagator:
\begin{equation}\label{2d}
D_{nm}(p) = D(p) \delta_{nm} = \frac{i}{p_0 -
  \frac{{\bf p}^2}{2 m} + i \epsilon}\; \delta_{nm}
\end{equation}

Ghost field propagator:
\begin{equation}
G_{ba}(p) = G(p) \delta_{ba} = \frac{-i}{{\bf p}^2}\;\delta_{ba}.
\end{equation}

\noindent
The vertices of the Feynman's diagrams are of five different types: 

Trilinear CS - matter field vertices ($p$ and $p^\prime$ are the momenta 
through the scalar lines at the vertex)
\begin{equation}
\Gamma^{a,0}_{nm}(p,p^\prime)= -g (T^a)_{nm}
\end{equation}
for the trilinear coupling involving  $A^0$, and 
\begin{equation}
\Gamma^{a,i}_{nm}(p,p^\prime)=-\frac{g}{2m} (T^a)_{nm}
(p+p^\prime)^i 
\end{equation}
for the coupling containing $A^i$, $i=1,2$.

Trilinear CS - ghost field vertex
\begin{equation}
\Gamma^{abc,i}(p,p^\prime)=-g f^{abc} p^{\prime i}.
\end{equation}

Quadrilinear CS matter field vertex
\begin{equation}
\Gamma^{ab,ij}_{nm}(p,p^\prime)=-\frac{i g^2}{2m} [T^a T^b+
T^b T^a]_{nm} g^{ij}.
\end{equation}

Trilinear CS field vertex
\begin{equation}
\Gamma^{abc,\mu\nu\lambda}(p,p^\prime)=
i g \Theta f^{abc} \varepsilon ^{\mu\nu\lambda}
\end{equation}

Quadrilinear matter field vertex
\begin{equation}
\Gamma(p,p^\prime)_{m^\prime n^\prime m n}=\frac{-i}{2}
C_{m^\prime n^\prime m n}
\end{equation}

Using these rules, the tree approximation to the direct scattering
amplitude corresponds to the graphs in Fig. \ref{fig1}$(a-b)$. In the
center of mass frame it  is given by
\begin{equation}\label{4}
{\cal M}(\theta)= -\frac{C}{2} - i \frac{2 \pi}{m} \Omega \;\cot(\theta/2)
\end{equation}
where $\theta$ is the scattering angle and $\Omega=\frac{-g^2}{2\pi
  \Theta} T^a \otimes T_a $. We use  a simplified notation
introduced in Ref. \cite{Bergman}, where isospin indices are omitted. 
Accordingly, if the incoming and outgoing particles have
isospin $(n,m)$ and $(n^\prime,m^\prime)$ the total scattering
amplitude for the process is given by
\begin{equation}\label{4a}
{\cal M}_{n^\prime m^\prime;nm}(\theta) + (\theta \rightarrow \pi+\theta, 
n^\prime\leftrightarrow m^\prime)
\end{equation}
where,
\begin{equation}\label{4b}
{\cal M}_{n^\prime m^\prime;nm}(\theta)= -\frac{C_{n^\prime m^\prime n m}}{2}+
i\frac{g^2}{\Theta} T^{a}_{n^\prime n}T^{a}_{m^\prime m} \; \cot(\theta/2)
\end{equation}

The one-loop graphs depicted in Fig. \ref{fig1}$(c-f)$ have been computed in 
\cite{Bergman}
using dimensional regularization. Here we just quote the  corresponding
results obtained by introducing a cutoff $\Lambda_{NR}$ in the spatial part
of the loop integrals. We get 

\begin{eqnarray}
{\cal M}_c(\theta)&=&\frac{m C^2}{16 \pi} \left[ \log
\left( \frac{\Lambda^2_{NR}}{{\bf p}^2}\right) + i \pi \right] \label{5a} \\
{\cal M}_d(\theta)&=&\frac{-\pi \Omega^2}{m} \left[ 2\; \log \mid 2\;
\sin (\theta/2) \mid + i \pi \right] \label{5b}\\
{\cal M}_e(\theta)&=&\frac{-\pi \Omega^2}{m} \left[ \log
\left( \frac{\Lambda^2_{NR}}{{\bf p}^2}\right)- 2\; \log\mid 2\;
\mbox{sin} (\theta/2) \mid\right]
\nonumber \\
& &+\frac{ g^2}{16m\Theta} \Omega  \left[ \log
\left( \frac{\Lambda^2_{NR}}{{\bf p}^2}\right)- 2\; \log\mid 2\;
\sin (\theta/2) \mid\right]\\
{\cal M}_f(\theta)&=&\frac{- g^2}{16m\Theta} \Omega  \left[ \log
\left( \frac{\Lambda^2_{NR}}{{\bf p}^2}\right)- 2\; \log\mid 2\;
\sin (\theta/2) \mid + 1\right] \; .
\label{5c}\end{eqnarray} 

\noindent
where ${\cal M}_i$ denotes the contribution coming from the graph $i$ in
Fig. \ref{fig1}. Note that the finite constant term in ${\cal M}_f$ can be
absorbed into a redefinition of $C$. Afterwards,
we see that our result agrees with Ref. \cite{Bergman} if the
dimensional regularization parameter $\epsilon$ and our cutoff $\Lambda_{NR}$
are related by
\begin{equation}\label{6}
\frac{1}{\epsilon}+ \ln\left(\frac{4 \pi \mu^2}{\Lambda^{2}_{NR}}\right ) -
\gamma =0,
\end{equation}
where $\gamma$ is the Euler constant. Adding the above results  
and disregarding the mentioned constant term, we get

\begin{equation}
{\cal M}_{1Loop}(\theta) = \frac{m}{16 \pi} 
\left( C^2- \frac{16\pi^2 \Omega^2}{m^2}\right) 
\left[ \log \left(\frac{\Lambda^2_{NR}}{{\bf p}^2}\right) + i \pi\right]\;.
\label{7}
\end{equation}

Thus, in complete analogy with what happens in  the Abelian case, 
we now choose $C^2=16 \pi^2 \Omega^2/m^2$
to restore the conformal
invariance of the tree approximation and reproduce the exact result 
\begin{equation}\label{7a}
{\cal M}(\theta) = -i\frac{2\pi}{m}  (\Omega \cot(\theta/2)- i |\Omega|),
\end{equation}
as also
derived by \cite{Bergman}.

\section{RELATIVISTIC THEORY}\label{section2} 
In the relativistic domain, the Feynman amplitudes are very intricate
for a general matrix $C$. Considerable simplification occurs however if
the gauge symmetry is taken as isospin $SU(2)$ group \cite{Kogan}. Thus, for
simplicity we restrict ourselves to the study of the $SU(2)$ case
described by the Lagrangian (\ref{1}) where a gauge fixing and ghost
terms must be added for a proper quantization.  

Of course, we choose to work in a strict Coulomb gauge as before. 
As described below, unless by obvious modifications the graphical 
representation of Feynman
amplitudes are the same as before. However, as indicated in Fig. \ref{fig2}, to
better clarify the matrix structure of the four scalar field vertices
we will sometimes use an auxiliary dotted line whose corresponding
propagator is just the identity. Concerning the free field
propagators, we have to use the matter field relativistic propagator
 \begin{equation}
\Delta_{nm} (p) = \Delta (p) \delta_{nm} = \frac{i}{p^2-
  m^2 + i \epsilon}\; \delta_{nm} \;  
\label{8}
\end{equation}
instead of (\ref{2d}) whereas the CS free propagator continues to be
given by (\ref{2c}). The new rules for the vertices are

Trilinear CS - matter field vertex
\begin{equation}
\Gamma^{a,\mu}_{nm}(p,p^\prime)= -g (T^a)_{nm} (p+p^{\prime})^\mu .
\end{equation} 

Quadrilinear CS matter field vertex
\begin{equation}
\Gamma^{ab,\mu\nu}_{nm}(p,p^\prime)=-i g^2 [T^a T^b+
T^b T^a]_{nm} g^{\mu\nu}
\end{equation}

Quadrilinear matter field vertices
\begin{equation}
\Gamma^{n^\prime m^\prime n m}_1(p,p^\prime) =\frac{-i\lambda_1}{2} 
{\it I}^{n^\prime n} {\it I}^{m^\prime m}
\end{equation}
for the vertex proportional to $\lambda_1$ ($I$ denotes the identity matrix in the isospin space). We have also
\begin{equation}
\Gamma^{n^\prime m^\prime n m}_2(p,p^\prime)
=\frac{-i\lambda_2}{2} (T^a)^{n^\prime n} (T^a)^{m^\prime m}
\end{equation} 
for the vertex proportional to $\lambda_2$. The CS - ghost field vertex
and the trilinear CS field vertex are, up the replacements of $f^{abc}$
by $\epsilon^{abc}$, the same as before.

Before embarking into the discussion of the scattering process and its
nonrelativistic limit we will examine the other one loop
superficially divergent amplitudes. The Coulomb gauge CS theory
without matter fields has been analyzed in Ref \cite{Ferrari} with the
conclusion that there are no radiative corrections to the Green
functions. For that reason we will restrict our study to graphs
arising from the coupling to the scalar matter field.

We begin by considering the matter self-energy contributions whose
nonvanishing contributions are in Figs. \ref{fig3}$(a-c)$. Because of the
specific form of the CS field propagator or the trace over the $SU(2)$
matrices, the other graphs are zero. We have,
\begin{eqnarray}
\Sigma^{nm}_{a} (p) &=& \delta^{nm} \Sigma_{a} (p)
\\
\Sigma^{nm}_{b} (p) &=& tr({\it I}) \delta^{nm} \Sigma_{a} (p)
\\
\Sigma^{nm}_{c} (p) &=& [T^aT_a]^{nm}  \Sigma_{a}
(p)_{\lambda_1\rightarrow \lambda_2}
\; ,
\label{9}\end{eqnarray}

\noindent
where, again the subscripts are in a strict correspondence with the diagrams
mentioned and
\begin{equation}
\Sigma_{a} (p)= -i\frac{\lambda_1}{2} \int \frac{d^3k}{(2\pi)^3}
\Delta (k) 
=\frac{\lambda_1}{2} \int^{\Lambda_0} \frac{d^3k}{(2\pi)^3} 
\frac{1}{k^2-m^2+i\epsilon}\; .
\label{10}
\end{equation}

\noindent
where $\Lambda_0$ is an ultraviolet cutoff in the spatial part of the above
integral. The integral is linearly divergent, but this divergence
could be absorbed by a mass renormalization.

Let us now look at the CS field polarization tensor. The  nonvanishing graphs
are shown in Figs. \ref{fig3}$(d-e)$. Up to a group
factor, they give the same result as in the Abelian case \cite{Gomes1}.
\begin{equation}
\Pi^{\mu\nu,ab}(q)=\frac{i g^2}{8\pi} tr(T^aT^b)(q^2 g^{\mu\nu} 
- q^\mu q^\nu) \Pi (q^2) \; ,
\label{11}
\end{equation}

\noindent
where
\begin{equation}
\Pi (q^2) = \int_0^1 dx \frac{1-4x+4 x^2}{[m^2- q^2 x(1-x)]^{(1/2)}}
\approx \frac{1}{3m} \left[1 + \frac{q^2}{20 m^2}\right] \; ,
\label{12}
\end{equation}
where the approximated result is valid for $m$ large compared with the 
momentum $q$. It is a purely relativistic quantum field theory contribution.

Consider now the corrections to the trilinear CS-matter field vertex
as shown in Fig. \ref{fig4}. Graphs \ref{fig4}$(e-h)$ vanish 
due to the oddness of the integrand. Although not so obvious, graph d also
vanishes. Since the integrand is $k_0$ independent, this result
follows by first regularizing the  $k^0$ part of the integral. More details
on this will be given later, when discussing the two body scattering
amplitude. 
Graph \ref{fig4}$a$ has the analytic expression
\begin{equation}
\Gamma^{\mu,b}_{(a)nm} (p,p^\prime) = [T^c T^b T_c]_{nm}
\Gamma^{\mu}_{(a)} (p,p^\prime)  
\label{13}
\end{equation}
where
\begin{equation}
\Gamma^{\mu}_{(a)} (p,p^\prime)= \frac{-g^3}{\Theta}
\int\frac{d^3k}{(2\pi)^3}
\left[ \frac{(2p-k)^\sigma
    \varepsilon_{\sigma\rho\lambda}\bar{k}^\lambda
 (2p^\prime -k)^\rho (p+p^\prime -k)^\mu}
{{\bf k}^2 [(p-k)^2 -m^2 +i\epsilon] [(p^\prime -k)^2 -m^2
  +i\epsilon]}\right]\; .
\label{14}
\end{equation}

\noindent
In the low momentum regime, this expression is easily integrated and we
get,
\begin{eqnarray}\label{15}
\Gamma^{0}_{(a)} (p,p^\prime)&=&\frac{ig^3}{4 \pi \Theta m}
\varepsilon_{ij} p^i q^j\; ,\\
\Gamma^{l}_{(a)} (p,p^\prime)&=&\frac{-ig^3}{8 \pi \Theta }
\left[ \varepsilon_{ij} \frac{p^i q^j}{m} 
\left( \frac{p^l+p^{\prime l}}{2m}\right) - 2 \varepsilon^{il} q_i 
\left( 1 + \frac{{\bf p}^2}{12 m^2} [2 + \cos (\theta)]\right) \right] \; ,
\label{16}
\end{eqnarray}

\noindent
where $q= p-p^\prime$. 

Graphs \ref{fig4}$b$ and \ref{fig4}$c$ have the expressions
\begin{eqnarray}
\Gamma^{\mu,a}_{(b)nm} (p,p^\prime)&=&[(T^{a}T^c+ T^c T^{a})T_c]_{nm} 
\Gamma^{\mu}_{(b)} (p,p^\prime) \nonumber \\
\Gamma^{\mu,a}_{(c)nm} (p,p^\prime)&=&[T_b (T^{a}T^b+T^b T^{a})]_{mn} 
\Gamma^{\mu}_{(c)} (p,p^\prime) \; ,
\label{17}
\end{eqnarray}
where
\begin{eqnarray}
 \Gamma^{\mu}_{(b)} (p,p^\prime) = \frac{g^3}{\Theta}\int\frac{d^3k}{(2\pi)^3} 
\left[ \frac{(2p-k)^\sigma \varepsilon_{\sigma\rho\lambda}
\bar{k}^\lambda  g^{\rho \mu}}
{{\bf k}^2 [(p-k)^2-m^2+i\epsilon]}\right] \\
 \Gamma^{\mu}_{(c)} (p,p^\prime) = \frac{g^3}{\Theta}\int\frac{d^3k}{(2\pi)^3} 
\left[ \frac{(2p^\prime-k)^\sigma \varepsilon_{\sigma\rho\lambda}
\bar{k}^\lambda  g^{\rho \mu}}
{{\bf k}^2 [(p^\prime-k)^2-m^2+i\epsilon]}\right] \; .
\label{18}
\end{eqnarray}

\noindent
Performing the integrals, we obtain
\begin{eqnarray}
 \Gamma^{0}_{(b)} (p,p^\prime)& =&\Gamma^{0}_{(c)} (p,p^\prime)\approx 0
\label{19}\\
 \Gamma^{l}_{(b)} (p,p^\prime) +\Gamma^{l}_{(c)} (p,p^\prime)&=&
 \frac{ig^3}{4 \pi \Theta} \varepsilon_{ij} g^{jl} q^i f({\bf p}^2,m^2)\; ,
\label{20}
\end{eqnarray}
where

\begin{equation}
f({\bf p}^2,m^2) = \frac{\sqrt{m^2+{\bf p}^2}}
{\sqrt{m^2} +\sqrt{m^2+{\bf p}^2}}
\approx \frac12 +\frac{{\bf p}^2}{8 m^2} \; .
\label{20a}
\end{equation} 

We postpone the computation of the graph \ref{fig4}$i$ till the
discussion of the two body scattering, to be done shortly. 

Besides the two body scattering amplitude, to conclude
our study of the one loop divergences we still have to look at the
corrections to the trilinear CS field vertex. There is just one graph
consisting of a closed loop of matter field propagators which, by power
counting, is logarithmically divergent (the other graphs cancel, as
shown in \cite{Ferrari}). However, as it is easily checked, it is in fact
convergent by symmetric integration. We will not   
explicitly compute  this diagram since, up to one loop, it does not contribute to
the scattering; we just want to remark that,
together with the vacuum polarization calculated before and the
one loop CS field four point function, it implies that a Yang-Mills kinetic
term 
\begin{equation}
\frac{1}{4} \frac{g^2}{24\pi m} tr[F^{\mu\nu} F_{\mu\nu}]
\label{21}
\end{equation}

\noindent
is induced in the effective low momentum Lagrangian.  
 So, up to this point, only a trivial
mass  renormalization of the matter field is necessary.

We are now ready to pursue our study of the two body scattering
process.  In our computations we will work in the center of mass frame
and shall retain terms up to order $|{\bf p} |^2/m^2\approx |{\bf
  p}|^4/\Lambda^4\approx \Lambda^4/m^4$, where ${\bf p}$ is the
momentum of the incoming particles and $\Lambda$ the intermediary
cutoff which separates the regions of low and high momenta in the
spatial part of the Feynman integrals. 

In the tree approximation, the contributing graphs are those shown in Fig. \ref{fig5}.
We get
 \begin{equation}
{\cal M}_{tree}(\theta)= - \frac{\lambda_{1}[{\it I}\otimes  {\it I}] + 
\lambda_2 [T^a\otimes T_a]}{2} - i (8 \pi ) \Omega\;
\omega_p \cot(\theta/2). \; 
\label{30}
\end{equation}   

Of course, we may adjust the coupling constants to correctly reproduce
the nonrelativistic result. Prior to that, however, we will examine
the one loop contributions which are listed in Figs. \ref{fig6}, \ref{fig7} and
\ref{fig8}. 

  Let us begin
by looking at the 2$^{nd}$ order (in $\lambda_1$ and $\lambda_2$) graphs.
The box diagrams, Figs. \ref{fig6}$(a1-a4)$, furnish
 \begin{equation}
{\cal M}_{\lambda^2(a)}(\theta)= \frac14 \left[-\lambda_1^2
 {\it I}\otimes {\it I}  - 2 \lambda_1 \lambda_2 T^a \otimes
T_a - \lambda_2^2 T^{a}T^b \otimes T_a T_b\right] I^{(1)} (\theta)
\label{31}
\end{equation}
where
\begin{equation}
 I^{(1)} (\theta)= -i\int \frac{d^3k}{(2\pi)^3} \Delta (k) \Delta
 (p_1+p_2-k)\; .
\label{32}
\end{equation} 

\noindent
We now detail the calculation of the above expression, to illustrate
the general procedure we shall follow in the evaluation of the
contributing graphs.
Firstly we integrate in $k^0$  getting,
\begin{eqnarray}
I^{(1)}(\theta) &=& \frac{1}{32\pi^2} \int_0^{\infty}  
d ({\bf k})^2 \int_0^{2\pi} d\alpha
\frac{1}{w_k} \frac{1}{{\bf p}^2 -{\bf k}^2 +i\epsilon}
\nonumber
\end{eqnarray} 
where $w_k=\sqrt{{\bf k^2} +m}$. The angular $\alpha$ integration is
trivial and gives $2\pi$. In order to facilitate the taking of the
nonrelativistic limit it is useful to introduce an auxiliary cutoff
$\Lambda$, satisfying $|{\bf p}|\ll \Lambda \ll m$, which separates
the integral in two regions of {\it low} ($0\leq {\bf k}^2\leq
\Lambda^2$) and {\it high} (${\bf k}^2\geq \Lambda^2$) loop momentum.
In the {\it low} part of the integral the integrand is expanded
in power of $1/m$ whereas in the {\it high} part 
it is  approximated by a Taylor expansion around ${\bf p}=0$.
We thus arrive at
\begin{eqnarray}
I^{(1)}(\theta)&=& I^{(1)}_{low}(\theta)+I^{(1)}_{high}(\theta) \, , \\
I^{(1)}_{low}(\theta) &=& \frac{1}{16\pi} \int_0^{\Lambda} 
d ({\bf k})^2\frac{1}{w_k} \frac{1}{{\bf p}^2 -{\bf k}^2 +i\epsilon}\\
&=&\frac{-1}{16 \pi m} \left\{ \left(1-\frac{{\bf p}^2}{2 m^2}\right)
\left[ \log \left(\frac{\Lambda^2}{{\bf p}^2}\right) + i \pi\right] -
\frac{{\bf p}^2}{\Lambda^2} -\frac{{\bf
    p}^4}{2\Lambda^4}-\frac{\Lambda^2}{2m^2}
+ \frac{3\Lambda^4}{16m^4} 
\right\} \; ,
\label{33}\\
 I^{(1)}_{high} (\theta)&=& \frac{1}{16\pi} \int_{\Lambda}^{\infty} 
d ({\bf k})^2\frac{1}{w_k} \frac{1}{{\bf p}^2 -{\bf k}^2 +i\epsilon}\\
&=& \frac{1}{16 \pi m} \left\{ \left(1-\frac{{\bf p}^2}{2 m^2}\right) 
 \log \left(\frac{\Lambda^2}{4m^2}\right) -\frac{{\bf p}^2}{2 m^2} 
-\frac{{\bf p}^2}{\Lambda^2} -\frac{{\bf p}^4}{2\Lambda^4}
-\frac{\Lambda^2}{2m^2}+ \frac{3\Lambda^4}{16m^4}
\right\} \; ,
\label{34}
\end{eqnarray}

\noindent
so that
\begin{equation}
 I^{(1)} (\theta)= 
\frac{-1}{16 \pi m} \left\{ \left(1-\frac{{\bf p}^2}{2 m^2}\right)
\left[ \log \left(\frac{4m^2}{{\bf p}^2}\right) + i \pi\right] +
\frac{{\bf p}^2}{2 m^2} \right\}
\; .
\label{35}
\end{equation}

\noindent
Notice that if we consider the {\it low} part of the above result and
reinterpret $\Lambda$ as the nonrelativistic $\Lambda_{NR}$ cutoff
then the leading contribution to (\ref{33}) is the same as the result
(\ref{5a}) from Sec \ref{section1}.  Here however, a counterterm is
automatically provided by the contribution of the {\it high} energy
part and, consequently, the final result (\ref{35}) is finite.
In the effective field theory program \cite{Weinberg} the high energy parts,
which are only polynomials in ${\bf p^2}$,
are associated to new interactions to be introduced in the nonrelativistic
Lagrangian.

Following the same steps described above, the sum of the remaining graphs, Figs. \ref{fig6}$(b-c)$, give 
\begin{eqnarray}
{\cal M}_{\lambda^2(bc)}(\theta)&=&
-\frac{1}{64\pi m}\left[  \left(-5\lambda_1^2 
+\frac{3}{2}\lambda_1\lambda_2
-\frac{3}{16} \lambda_2^2 \right) {\it I}\otimes {\it I}
\right.\nonumber \\
&+&\left.
\left(  - 4 \lambda_1 \lambda_2 +\frac12
\lambda_2^2\right) T^a \otimes T^a\right] \left( 2 -\frac{{\bf p}^2}{3
m^2}\right)  
\nonumber \\ 
&-&\frac{1}{64\pi m}\left[  \left(-3\lambda_1^2 
+\frac{3}{2}\lambda_1\lambda_2
+\frac{3}{16} \lambda_2^2 \right) {\it I}\otimes {\it I}
- \frac12\lambda_2^2  T^a \otimes T^a\right]  \frac{{\bf p}^2}{3
m^2}\cos \theta   
\label{36}\end{eqnarray}

As expected, the above result come essentially from the high energy
part of the corresponding integrals. Let us now examine the graphs
involving the CS field which are listed in Fig. \ref{fig7}. The graphs in the
first row, the direct box and twisted box diagrams are given
respectively by

\begin{eqnarray}
{\cal M}_{g^4(a)}(\theta)&=& -i g^4 [T^aT^c \otimes T_a T_c]
\int \frac{d^3k}{(2 \pi)^3} \left\{ \Delta (k) \Delta (p_1+p_2-k)(p_1+k)^\mu
\right.\nonumber \\
&&\left.  D_{\sigma\mu} (k-p_1) (2 p_2+p_1 -k)^\sigma
[(p_3+k)^\nu D_{\nu\rho} (k-p_3) (p_2+p_1+p_4 -k)^\rho] \right\},
\nonumber \\
\label{37}
{\cal M}_{g^4(b)}(\theta)&=& -i g^4 [T^aT^c \otimes T_c T_a]
\int \frac{d^3k}{(2 \pi)^3} \left\{ \Delta (k) \Delta (k+p_2-p_3)(p_1+k)^\mu
\right. \nonumber \\
& &\left. 
 D_{\mu\sigma} (p_1-k) (k+p_2-p_3 +p_4 )^\sigma
[(2 p_2-p_3+k)^\nu D_{\nu\rho} (p_3-k) (k+p_3)^\rho] \right\}.
\nonumber \\
& &
\label{38}
\end{eqnarray} 
Performing the integrals one finds that the {it low} energy part, after
the reinterpretation of the intermediary cutoff, has a logarithmic
divergence which is canceled by the contribution from the high energy
part. The final result for the sum of the these graphs is 
\begin{eqnarray}
{\cal M}_{g^4(ab)}(\theta)&=&\frac{-g^4m}{2 \pi \Theta^2} 
[ T^a T^c \otimes T_aT_c]
\left\{2\left(1 +\frac{{\bf p}^2}{2m^2}\right) \left[\log[2(1-\cos\theta)]+i\pi\right]+
\right.\nonumber \\
&-&2 \cos\theta \left. \left[\log[2(1-\cos\theta)] -\log\left(\frac{4 m^2}{{\bf p}^2}\right) \right]
\frac{{\bf p}^2}{m^2} -\frac{{\bf p}^2}{m^2} (1+\cos \theta)
\right\}+\nonumber \\
&+&\frac{g^4m}{4 \pi \Theta^2}  \left[ T^a \otimes T_a\right]
\left(2 (1+\cos \theta ) \frac{{\bf p}^2}{m^2} \right)
\label{39}
\end{eqnarray}

The next set of diagrams in Fig. \ref{fig7} needs special care since
they contain a genuine ultraviolet divergence of the relativistic
theory. In fact they may have divergences in both $k_0$ and $\bf k$
parts of the loop momentum integral. We found that they are naturally
grouped in two sets which have distinct properties. We shall discuss
each of them separately.

The graphs of the first set, Figs. \ref{fig7}$c$ and \ref{fig7}$d$, are potentially more dangerous for, because of the form taken by the
CS propagator, the $k^0$ integration is not well defined. We found that, as
suggested in \cite{Ferrari}, this difficulty can be circumvented by
first regularizing  the $k_0$ integral. For practical purpose we
will  introduce an additional cutoff so that
\begin{equation}\label{41} 
\int dk_0 f(k_0) \rightarrow \int_{-L}^{L} f(k_0),
\end{equation}

\noindent
but the final result  actually does not depend on the regularization one
chooses.  It turns out that the spatial integral multiplying the $k_0$
integral vanishes. To see how this happens, consider the graph $c$
whose analytic expression is
\begin{eqnarray}
{\cal M}_{g^4(c)}(\theta)&=& -g^4\Theta [ \varepsilon_{bac} T^b
\otimes T^c T^a] D_{\sigma^\prime \sigma} (q)
\varepsilon^{\sigma^\prime \mu^\prime \nu^\prime} (p_2+p_4)^\sigma
\nonumber \\
& &\int \frac{d^3k}{(2\pi)^3} [(k+p_3)^\nu \Delta (k) (k+p_1)^\mu ]
D_{\mu^\prime \mu} (k-p_1) D_{\nu\nu^\prime}(k-p_3)\;\label{42}
\end{eqnarray}

\noindent
After some arrangement, the above expression can be rewritten as
\begin{equation}
{\cal M}_{g^4(c)}(\theta)= \frac{-2 g^4}{\Theta^2}
 [ \varepsilon_{bac} T^b \otimes T^c T^a]
\int \frac{d^2{\bf k}}{(2\pi)^2}
 \frac{[{\bf q} \wedge {\bf k} -{\bf p}_1 \wedge {\bf p}_3]}
{[({\bf k}-{\bf p}_1)^2 ] 
[({\bf k}-{\bf p}_3)^2] [{\bf q}^2]} T_0 \; ,
\label{43}
\end{equation}
where
\begin{eqnarray}
T_0= \int_{-L}^{L} \frac{d k^0}{(2\pi)}
\frac{ 2 w_q (k^0 + w_q) [ {\bf q} \wedge {\bf k}] 
+ (k^0 + w_q)^2 [ {\bf p}_1 \wedge {\bf p}_3] }{k_0^2 - w_k^2
+ i\epsilon}\; .
\label{44}
\end{eqnarray}

For $L$ large enough, we obtain
\begin{equation}
T_0= [ {\bf q} \wedge {\bf k}] \frac{w_p^2}{w_k} 
+  [ {\bf p}_1 \wedge {\bf p}_3] \frac{w_p^2 + w_k^2}{2 w_k} + 
\frac{L}{\pi}[ {\bf p}_1 \wedge {\bf p}_3]
\label{45}
\end{equation}
Notice now that the spatial integral multiplying the divergent piece in
$T_0$ is 
\begin{equation}
\int \frac{d^2{\bf k}}{(2\pi)^2}
 \frac{({\bf k-p_1}) \wedge ({\bf k-p_3})}
{[({\bf k}-{\bf p}_1)^2 ] 
[({\bf k}-{\bf p}_3)^2] }=0 \; ,\label{46} 
\end{equation}
so that no counterterm will be needed if we agree to eliminate the cutoff
$L$ only at the end of the calculation. Proceeding in this way  and making the
nonrelativistic approximation we arrive at
\begin{eqnarray}
{\cal M}_{g^4(cd)}(\theta)&=&-\frac{g^4m}{4 \pi \Theta^2} 
\left[  T^a \otimes T_a\right]
\left\{2 + \left[\frac{11}{2}+ 3\cos\theta \right] \frac{{\bf p}^2}{m^2} +
\right. \nonumber \\
&+&\left.
\left(2 + [1-2 \cos \theta]  \frac{{\bf p}^2}{m^2}\right)
\left[ \log\left(\frac{4m^2}{{\bf p}^2}\right)
 -  \left[\log[2(1-\cos\theta)]\right] \right] 
\right\} \; .
\label{47}
\end{eqnarray}
for the sum of the graphs \ref{fig7}$c$ and \ref{fig7}$d$. 

Similarly to the calculation for the
\ref{fig7}$c$ graph, each one of the graphs \ref{fig7}$e$-\ref{fig7}$g$ presents a linear divergence in
L. This divergence is however eliminated when we sum the contributions
so that the final result is:

\begin{eqnarray}
{\cal M}_{g^4(efg)}(\theta)&=&\frac{-g^4m}{2 \pi \Theta^2} 
\left[T^a T^c \otimes T_a T_c\right]
\left\{-2 + (2+ \cos\theta ) \frac{{\bf p}^2}{m^2} +
\right. \nonumber \\
&+&\left. \left(2 + [1-2 \cos \theta]  \frac{{\bf p}^2}{m^2}\right)
\left[ \log\left(\frac{4 m^2}{{\bf p}^2}\right)
 -  \left[\log[2(1-\cos\theta)]\right] \right]
\right\}\nonumber \\
&+&\frac{g^4m}{4 \pi \Theta^2} 
\left[  T^a \otimes T_a\right]
\left\{-2 + (2+ \cos\theta ) \frac{{\bf p}^2}{m^2} + \left(2 + [1-2 \cos \theta]  \frac{{\bf p}^2}{m^2}\right)
\right. \nonumber \\
&\times&\left. 
\left[ \log\left(\frac{4 m^2}{{\bf p}^2}\right)
 -  \left[\log[2(1-\cos\theta)]\right] \right]
\right\} - \frac{3 g^4 m}{16 \pi \Theta^2} [I\otimes I] \frac{\Lambda_0}{m}
\label{48}
\end{eqnarray}

\noindent
where $\Lambda_0$ in an ultraviolet cutoff introduced in the spatial
part of the integral.
The divergent term proportional to this cutoff can be removed
by a redefinition of the coupling constant $\lambda_1$. After this, the
final result for the sum of the graphs in Fig. \ref{fig7} is
\begin{eqnarray}
{\cal M}_{g^4}(\theta)&=&\frac{-g^4m}{2 \pi \Theta^2} 
\left[ T^a T^c\otimes T_aT_c\right]
\left\{2 \left(1+\frac{{\bf p}^2}{2m^2}\right) 
\left[\log\left(\frac{4 m^2}{{\bf p}^2}\right) +i\pi\right]
 + \frac{{\bf p}^2}{m^2} \right\}
 \nonumber \\
&-&\frac{g^4m}{4 \pi \Theta^2} \left[T^a \otimes
  T_a\right]\frac32\frac{{\bf p}^2}{m^2}
\; .
\label{40}\end{eqnarray}

We still have to incorporate to our calculation the contributions of graphs with vacuum polarization and vertex corrections. They
do not exist in the nonrelativistic theory of the previous section and their
contribution come entirely from the {\it high} energy part. Using Eq. (\ref{11}), the insertions of
vacuum
polarization graphs give
\begin{equation}
{\cal M}_P (\theta)= \frac{g^4m}{4 \pi \Theta^2} [T^a \otimes T_a]
\left\{ \frac13 + \frac{7}{15}\frac{{\bf p}^2}{m^2}
 + \frac15 \frac{{\bf p}^2}{m^2} \cos \theta \right\}\; ,
\label{49}
\end{equation}

\noindent
whereas, using Eqs. \ref{15}, \ref{16} and \ref{20}, the vertex
corrections produce
\begin{equation}
{\cal M}_V (\theta)= -\frac{g^4m}{4 \pi \Theta^2} [T^a \otimes T_a]
\left\{\frac16  
    \frac{{\bf p}^2}{m^2} + 
\frac13 \frac{{\bf p}^2}{m^2} \cos \theta \right\}.
\label{50}
\end{equation}  

At last, there are some contributions from graphs that admix the quadrilinear
scalar vertex and the CS- matter field vertex.  They are shown  in Fig. \ref{fig8} and give the result 
\begin{equation}
{\cal M}_{\lambda g^2} (\theta)= 
-\frac{i g^2}{8 \pi\Theta} 
\left[ \frac34 \left(\lambda_1+ \frac{\lambda_2}{4}
  \right) {\it I}\otimes {\it I} 
+ \left( \lambda_1 + \frac34\lambda_2 \right)  T^a \otimes T_a \right] 
\frac{{\bf p}^2}{m^2}\sin \theta .
\label{51}
\end{equation} 
 In the Abelian situation the corresponding amplitude is canceled by
its exchanged particle partner. Here, because of the non-Abelian structure,
even after symmetrization the result is nonvanishing. 

Our one loop calculation is now completed. Collecting all the results above,
the total one loop amplitude (without symmetrization) is given by
\begin{eqnarray}
{\cal M}_{1Loop}&=& \frac{-1}{64\pi m} 
\left[  \left(-\lambda_1^2 -\frac{3}{16}
\lambda_2^2 \right) {\it I}\otimes {\it I} +
\left( -\frac12 \lambda_2^2 - 2 \lambda_1 \lambda_2\right) T^a \otimes
T_a\right]  \nonumber \\
& &\left\{ \left(1-\frac{{\bf p}^2}{2 m^2}\right)
\left[ \log \left(\frac{4m^2}{{\bf p}^2}\right) + i \pi\right] +
\frac{{\bf p}^2}{2 m^2} \right\}
\nonumber \\ 
&-&\frac{1}{64\pi m}\left[  \left(-5\lambda_1^2 
+\frac32\lambda_1\lambda_2
-\frac{3}{16} \lambda_2^2 \right) {\it I}\otimes {\it I}
+\left(  - 4 \lambda_1 \lambda_2 +\frac12
\lambda_2^2\right) T^a \otimes T_a\right] \left( -\frac{{\bf p}^2}{3
m^2}\right)  
\nonumber \\ 
&-&\frac{1}{64\pi m}\left[  \left(-3\lambda_1^2 
+\frac32\lambda_1\lambda_2
+\frac{3}{16} \lambda_2^2 \right) {\it I}\otimes {\it I}
-\left( \frac12\lambda_2^2 \right) 
T^a \otimes T_a\right]  \frac{{\bf p}^2}{3
m^2}\cos \theta   
\nonumber \\
&-&\frac{g^4m}{2 \pi \Theta^2} 
\left[T^a T^c\otimes T_a T_c\right]
\left\{2 \left(1+\frac{{\bf p}^2}{2m^2}\right) 
\left[\log\left(\frac{4 m^2}{{\bf p}^2}\right) +i\pi\right]
 + \frac{{\bf p}^2}{m^2} \right\}
 \nonumber \\
&-&\frac{g^4m}{4 \pi \Theta^2} \left[ T^a \otimes
  T_a\right]
\left\{ \frac{18}{15}\frac{{\bf p}^2}{m^2} + \frac{2}{15}
 \frac{{\bf p}^2}{m^2} \cos \theta\right\}
\nonumber \\
&-&\frac{i g^2}{8 \pi\Theta} 
\left[ \frac34 \left(\lambda_1+ \frac{\lambda_2}{4}
  \right) 
{\it I}  \otimes {\it I} + \left( \lambda_1 - \frac34\lambda_2\right) 
 T^a \otimes T_a \right] \frac{{\bf p}^2}{m^2} \sin \theta
\nonumber\\
& & 
\label{52}
\end{eqnarray}

We are now in a position to compare the results of this section with
the ones stated in the preceding section. Before doing that we will need to
adjust some normalization factors. These come from the normalization
of the relativistic one particle states, $\langle {\bf p}^\prime\mid {\bf p}\rangle= 2 \omega_p
\delta( {\bf p}^\prime- {\bf p})$, whereas the nonrelativistic theory does not
have the $2 w_p$ factor. Another factor to take into consideration is
$f=\sqrt{\omega_p/m}$  which comes from the different expressions used for the
relativistic and nonrelativistic velocities. Altogether, we need to multiply
the relativistic expression by the kinematic factor
\begin{equation} 
f \left(\frac{1}{\sqrt{2 \omega_p}}\right)^4= \frac{1}{4m^2}
\left[1-\frac{3{\bf p}^2}{4m^2}+ ...\right]\; ,
\label{53}
\end{equation}

\noindent
Thus, to leading order in ${\bf p}/m$, we have
\begin{eqnarray}
{\cal M}^{Dom} (\theta)&=& - \frac12\left(\frac{\lambda_{eff}}{4m^2}\right) 
- i \frac{2 \pi}{ m} \Omega  \cot(\theta/2) 
+\frac{m}{16 \pi} \left[\left(\frac{\lambda_{eff}}{4 m^2}\right)^2 
- \frac{16 \pi^2}{m^2}
  \Omega^2\right]
 \left\{ \log \left[\frac{4m^2}{{\bf p}^2}\right] + i \pi\right\}\; ,
\nonumber \\
& &
\label{54}\end{eqnarray}
where $\lambda_{eff}= \lambda_1 [{\it I}\otimes {\it I}] + \lambda_2
[T^a\otimes T_a]$. As remarked after Eq. (\ref{35}) and it is explicit
in the equations (\ref{33}) and (\ref{34}) the low energy part of this
formula coincide with the nonrelativistic result, after the
identification of the intermediary with the nonrelativistic cutoff. In
our calculation however, the {\it high} part provides the necessary
counterterm to the {\it low} part and the final result becomes
automatically finite.

To restore the conformal invariance of the tree approximation,
eliminating the $\log$ term and obtaining the same result as in the
expansion of the exact amplitude one must choose $\lambda_1=0$ and
$\lambda_2= 8mg^2/(|\Theta|)$.  At these values of the renormalized
quartic self interaction there are subdominant terms in (\ref{52}) 
that survive, namely,

\begin{eqnarray}
{\cal M}^{Sub} &=&
\frac{i\pi}{2 m}\left[ \Omega \cot (\theta/2) - 
i\frac34 |\Omega| \right] \frac{{\bf p}^2}{m^2}
\nonumber \\
 &-&\frac{g^4 }{4\pi m\Theta^2}
 \left[\frac{3}{16} {\it I}\otimes {\it I} 
-\frac12 T^{a}\otimes T_a \right]\frac{{\bf p}^2}{m^2}
\left( \log \left[\frac{4m^2}{{\bf p}^2} \right] +i\pi \right)
\nonumber \\
&-&\frac{g^4 }{4\pi m\Theta^2} \left\{
 \left[\frac{1}{16} {\it I}\otimes {\it I} 
+\frac{2}{15} T^{a}\otimes T_a \right]\frac{{\bf p}^2}{m^2} 
+  \left[\frac{1}{16} {\it I}\otimes {\it I} 
-\frac{2}{15} T^{a}\otimes T_a \right]\frac{{\bf p}^2}{m^2}\cos \theta\right\}
\nonumber \\
& \pm &\frac{ig^4}{4\pi m \Theta |\Theta |} 
 \left[\frac{3}{16} {\it I}\otimes {\it I} 
-\frac{3}{4} T^{a}\otimes T_a \right]\frac{{\bf p}^2}{m^2}\sin \theta
\end{eqnarray}

\noindent
and represent relativistic corrections to the non-Abelian AB
scattering. These terms break conformal invariance which is therefore
only a property of the leading approximation. The first row of the
above formula are corrections of the tree level and are partially
of kinematical rise and partially due to the energy dependence of the
relativistic amplitude Eq. (\ref{30}). The other terms, proportional to $g^4$
are absent in the nonrelativistic Aharonov-Bohm scattering which contains
only odd powers of $g^2$ \cite{Bergman}. In an effective low momentum
Lagrangian they would correspond to derivative quartic self-couplings
of the matter field $\phi$.
The subleading corrections
change also the nature of the effective AB potential: because
of vacuum polarization, this potential is not strictly localized at the
origin  and the AB effect, considered as the scattering by an impenetrable
flux tube, only exists in a quantum mechanical, first quantized
level.

\section*{acknowledgments}

This work was partially supported by Conselho Nacional de
Desenvolvimento Cient\'{\i}fico e Tecnol\'ogico (CNPq) and Coordena\c
c\~ao de Aperfei\c coamento de Pessoal de N\'{\i}vel Superior (Capes).

\begin{figure} 
\caption{Graphs contributing to the nonrelativistic scattering}
\label{fig1} 
\end{figure} 
\begin{figure}
\caption{Alternative graphical representation for the four scalar field 
vertex} \label{fig2} 
\end{figure} 
\begin{figure}
\caption{One loop contributions to the matter field and CS self-energy} 
\label{fig3} 
\end{figure} 
\begin{figure}
\caption{Corrections to the CS-matter field trilinear vertex} 
\label{fig4}
\end{figure} 
\begin{figure} 
\caption{Tree approximation  to the scattering} 
\label{fig5}
\end{figure} 
\begin{figure} 
\caption{$\lambda$'s second order contributions to the scattering} 
\label{fig6}
\end{figure}
\begin{figure} 
\caption{Graphs contributing to the scattering having only CS-matter field vertices} 
\label{fig7}
\end{figure}
\begin{figure} 
\caption{Graphs admixing the quadrilinear scalar and trilinear CS-matter field
vertices} 
\label{fig8}
\end{figure} 
\end{document}